\documentclass[doublecol]{epl2}
\usepackage{graphicx}
\usepackage{dcolumn}
\usepackage{bm}
\usepackage{amsmath}
\usepackage{color}
\usepackage{epsfig}
\usepackage{amssymb,amsmath}
\usepackage{eurosym}

\title{Dynamics of the wakefield of a multi-petawatt, femtosecond 
laser pulse \textcolor{blue}{in a configuration with ultrarelativistic electrons}}
\shorttitle{Ultrarelativistic wakefield}

\author{D. Jovanovi\'c\inst{1} \and R. Fedele\inst{2,3} \and M. Beli\'c\inst{4}}
\shortauthor{D. Jovanovi\'c \etal}

\institute{
  \inst{1} Institute of Physics, University of Belgrade, Pregrevica 118, 11080 Belgrade (Zemun), Serbia\\
  \inst{2} Dipartimento di Fisica, Universit\`{a} di Napoli "Federico II", Complesso Universitario di M.S. Angelo,
Napoli, Italy\\
  \inst{3} INFN Sezione di Napoli, Napoli, Italy\\
  \inst{4} Texas A\&M University at Qatar, P.O. Box 23874 Doha, Qatar
}
\abstract{The wake field excitation in an unmagnetized plasma by a multi-petawatt, femtosecond, pancake-shaped laser pulse is described both analytically and numerically in the \textcolor{blue}{regime with ultrarelativistic electron jitter velocities,} \textcolor{blue}{when the plasma electrons are almost expelled from the pulse region}. This is done, for the first time, in fluid theory. A novel mathematical model is devised \textcolor{blue}{that does not break down for very intense pump strengths, in contrast to the standard approach that uses the laser field envelope and the ponderomotive guiding center averaging}. This is accomplished by employing a three-timescale description, with the intermediate scale associated with the nonlinear phase of the electromagnetic wave and with the bending of its wave front. The evolution of the pulse and of its electrostatic wake are studied by the numerical solution in a two-dimensional geometry, with the spot diameter $\gtrsim 100 \, \mu$m. \textcolor{blue}{It reveals that the optimum initial pulse length needs to be somewhat bigger than $\gtrsim 1 \, \mu$m (1-2 oscillations), as suggested by simple analytical local estimates, because the nonlocal plasma response tends to stretch very short pulses.}}

\begin{document}

\maketitle

\section{Introduction}\label{Introductory}

Tajima and Dawson proposed \cite{LWF2} in 1979 to accelerate charged particles by large amplitude electron density waves, propagating through underdense plasma in the wake of an intense laser pulse. Plasma wakes can sustain electrostatic fields of several GV/cm, $10^3$ times above the electric breakdown in conventional accelerators, enabling the construction of low cost, miniature laser-plasma accelerators (LPAs)
\cite{2009RvMP...81.1229E}.
Most powerful LPA systems at present time, or planned for the near future \cite{2011PhyU...54....9K}, include the Nd:Glass lasers with the wavelength $\lambda = 1.06 \, \mu{\rm m}$, pulse duration $T = 300-500 \, {\rm fs}$ and power $P \lesssim 1 \, {\rm PW}$, and the facilities using Ti:Sapphire technology, $\lambda = 0.65-1.1 \, \mu{\rm m}$, having shorter pulses, $T =25-60 \,{\rm fs}$, and the power $P = 0.1-1 \, {\rm PW}$. 
So far, the maximum electron beam energy achieved in LPAs is $\gtrsim 1 \, {\rm GeV}$ \cite{2013NatCo...4E1988W}.
Fundamental limitation is set by the pump depletion. To produce a $10 \, {\rm GeV}$ electron bunch with a charge of $1 \, {\rm nC}$, holding $10 \, {\rm J}$ of kinetic energy, with a laser to particle beam efficiency $1 - 10$\%, laser energy of $100 - 1000 \, {\rm J}$ is needed, i.e. $P = 40-400 \, {\rm PW}$, if the pulse duration is $\sim 25 \, {\rm fs}$.

\textcolor{blue}{In the terminology introduced in Ref. \cite{moj_EPJD},} \textcolor{blue}{an LPA is said to be operating in the moderate (MIR) or in the strong intensity regime (SIR) when} \textcolor{blue}{the electron quiver motion is {\em mildly relativistic}, $p_{\bot_0} \lesssim m_0 c$ or {\em ultrarelativistic}, $p_{\bot_0}\gg m_0 c$, respectively (here $p_{\bot_0} = e E_{\bot_0}/\omega$ and $\omega$ and  $E_{\bot_0}$ are the angular frequency and the amplitude of the laser electric field). Simple scaling \cite{moj_EPJD} reveals that in the \textcolor{blue}{SIR} the electron density perturbation is comparable to the plasma density, i.e. that the ponderomotive force entirely routs plasma electrons from the pulse area and leaves a wake of immobile, positively charged ions. For a spheroidal laser pulse, whose length and width are comparable to the plasma length, $L_\Vert \sim L_\bot \sim 2\pi c/\omega_{pe}$, the threshold for the complete expulsion of the electrons was estimated to be $p_{\bot_0} \geq 4 \, m_0 c$ \cite{2006PhPl...13e6709L}. The phenomenological theory \cite{2004PhPl...11.5256K} found that such plasma cavity, or \textit{bubble}, develops
when the nonlinearities are sufficiently strong to produce a plasma wave breaking after the first oscillation. 3D PIC
simulations \cite{2002ApPhB..74..355P} confirmed the existence of the \textit{bubble}
regime in LPA and showed that a bubble can trap background electrons and accelerate them, with a monoenergetic spectrum.
\textcolor{blue}{Spheroidal bubbles are
inherently electromagnetic} \cite{2009RvMP...81.1229E}, since the wake is encircled by a sheath of relativistic electrons (return current) exerting a Lorentz force on  electrons. The balance of the Lorentz, Coulomb and ponderomotive forces determines the size of a 3D bubble.
For self-similar 3D pulses, the optimum wake generation \cite{2005PhPl...12d3109G,PukhovBubbles} occurs for the pulse length  $L_\Vert \leq d_s/2$ and the laser spot diameter $d_s = (2 c/\omega_{pe}) \sqrt{e E_{\bot_0}/\omega m_0 c}$. }

Writing the norm for the \textcolor{blue}{MIR  (mildly relativistic) and SIR (ultrarelativistic regime)} in terms of laser intensity $I = w c$, where $w$ is the energy density of the e.m. (electromagnetic) wave, $w = \epsilon_0 E_{\bot_0}^2/2$, we note that 
lasers with $\lambda \lesssim 1 \, \mu{\rm m}$ have $I \sim (\epsilon_0 \, c/2) (2 \pi \, m_0 c^2/e \lambda)^2
\sim 10^{18} \,{\rm W/cm}^2$ in the \textcolor{blue}{MIR} and $I \gg 10^{18} \, {\rm W/cm}^3$ in the \textcolor{blue}{SIR}. The diameter $d_s$ of the laser spot is estimated from $I = 4 P/\pi d_s^2$, which in the \textcolor{blue}{SIR} yields $d_s \ll 305 \, \lambda$ $\sqrt{P/(10^{15} \, {\rm W})}$. Thus, a $100 \, {\rm TW}$ class Ti:Saphire laser needs to be focussed to a diameter comparable to the pulse length,  $d_s \ll 60 \, \mu{\rm m}$.

In this letter, a theoretical study of the ultrarelativistic regime in the laser wake field generation is for the first time carried out in fluid theory. We consider the propagation, in a  cold unmagnetized plasma, of a pancake-shaped, ultra-short laser pulse, with the energy $ \geq 100 \, {\rm J}$ and the intensity $I \sim 10^{20} \, {\rm W/cm}^2$. A multi-petawatt pulse \textcolor{blue}{reaches an ultrarelativistic intensity that strongly perturbs the electron density (causing almost complete expulsion)} even if focussed to a large spot, $d_s \sim 100 \, \mu{\rm m} \ll 305 \, \lambda $ $ \sqrt{P/(10^{15} \, {\rm W})} \sim 1000 \, \mu{\rm m}$, much bigger than the length of a $25 \, {\rm fs}$ pulse ($L_\Vert \sim 7.5 \,\mu{\rm m}$). \textcolor{blue}{Some authors argue that a "pancake" shape is beneficial for LPA \cite{2005PhPl...12c3101G}, because} a tight laser spot $d_s\sim 10 \, \mu{\rm m}$ gives an acceleration length of only a few mm (estimated as twice the Rayleigh length), restricting the electrons' energy gain. \textcolor{blue}{Although a suitably preformed plasma and the nonlinear self-guiding may enable a tightly focussed laser pulse to propagate well beyond two-three Rayleigh lengths \cite{PukhovBubbles,1997PhRvL..78..879B,1997IJQE...33.1879E}, its strong radial electric field expels most electrons, permitting only a few to be trapped and accelerated by a 3D potential \cite{2005PhPl...12c3101G}.
}

\textcolor{blue}{
A \textcolor{blue}{SIR} involves vastly different scalings in the core and at the pulse edges. The core is almost devoid of electrons and the e.m. pulse practically propagates in vacuum, featuring linear properties. The nonlinear self-organization \cite{2005PhPl...12c3101G,moj_EPJD} occurs at the edges, which are in the} \textcolor{blue}{MIR.} For laser intensities $I \sim 10^{20} \, {\rm W/cm}^2$ we employ new model equations that describe both the \textcolor{blue}{SIR} core and the \textcolor{blue}{MIR} edges. \textcolor{blue}{This is not possible in the classical two-timescale description of a slowly varying amplitude of the laser pulse}. We develop a three-timescale description, with an intermediate timescale associated with the intensity-dependent phase of the e.m. pulse. At SIR intensities, the nonlinear phase is resolved within the (\textcolor{blue}{Wentzel-Kramers-Brillouin}) approximation. The phase introduces new nonlinear terms in the wave equation that suppress, \textcolor{blue}{in the core}, both the nonlocal nonlinearity and the dispersion of the e.m. wave. \textcolor{blue}{Our 2D numerical result reveals that, as} \textcolor{blue}{the core of the pulse runs at a higher group velocity than the leading edge, an initial steepening of the pulse's front edge takes place, which is known to occur in the absence of nonlocality (i.e. spatial dispersion) in the plasma response \cite{2001AIPC..569..214E}. Soon, the latter  produce an effective mixing of the rapid core of the pulse with its slow front edge, pushing it forward and producing a frontward stretching of the laser pulse. \textcolor{blue}{Remarkably, the stretched pulse propagates in the plasma several mm, consistent with the results of self-injection experiments} \cite{Gizzi2013}. }

\section{Mathematical model}\label{model}

\textcolor{blue}{Due to their extreme complexity, analytic studies of the laser-plasma interaction with intensities suitable for LPA have been attempted only for quasi 1D, pancake-shaped pulses, using the "quasistatic" approximation and in a cold-fluid description, see the classical papers \cite{sprangle,mahajan,Shukla,2005PhPl...12c3101G} and references therein. Recently, in the \textit{mildly relativistic regime}, the evolution of the plasma wake and of the laser pulse (depletion, frequency redshifting) was satisfactorily described using a reduced wave equation and a quasistatic plasma response \cite{2011PhRvL.106m5002S}, with a good agreement with full Maxwell-fluid results. Such fluid calculations provided an additional insight into purely particle phenomena, e.g. by establishing the appropriate thresholds for the electron trapping and wave breaking, and showed that the electron dephasing (rather than laser depletion) limits the LPA's energy gain. Following these works, considering an unmagnetized plasma, assuming $\nabla_\bot \ll \partial/\partial z$, and taking that the solution is slowly varying in the frame that moves with the velocity $u\,\vec{e}_z$, we have derived our system \textit{wave equation} + \textit{Poisson's equation}, (\ref{waveqenv0}), (\ref{poissons0}). Their derivation is given in \cite{moj_EPJD} and we note that Eqs. (\ref{waveqenv0}) and (\ref{poissons0}) are valid also for {\em ultrarelativistic} electrons, $p_{\bot_0}\gg m_0 c$.} \textcolor{blue}{Being affected by the return electron current, a wake is innately electromagnetic, but for sufficiently broad pulses, both pancake-shaped \cite{moj_EPJD,2011PhRvL.106m5002S} and spherical \cite{2010PhPl...17e4704K}, the electromagnetic effects are weak and we consider the wake as purely electrostatic.
}

Using the normalizations
$\vec{p} \to {\vec p}/{m_0 c}$,
$\vec{v} \to {\vec v}/{c}$,
$\phi \to {-e\phi}/{m_0 c^2}$,
$\vec{A} \to {-e\vec A}/{m_0 c}$,
$u \to {u}/{c}$,
$t \to \omega_{pe} t$,
${\vec r} \to ({\omega_{pe}}/{c}) (\vec r - \vec e_z\, u t)$,
our basic equations take the form
\begin{eqnarray}
&\left[\frac{\partial^2}{\partial t^2} - 2
u\,\,\frac{\partial^2}{\partial t\,\, \partial z} -
\left(1-u^2\right)\frac{\partial^2}{\partial z^2} - \nabla_\bot^2
+ \frac{1}{1-\phi}\right]\vec A_\bot
\label{waveqenv0}\nonumber\\
& =
-\left(\frac{\partial}{\partial t} - u\,\,
\frac{\partial}{\partial z}\right)\nabla_\bot\phi,
\end{eqnarray}
{{\small
\begin{equation}
\label{poissons0}
\frac{\partial^2\phi}{\partial z^2} = \frac{\left(\phi - 1\right)^2 - 1 - {\vec{A}_\bot}^{\, 2}}{2 \left(\phi - 1\right)^2}.
\end{equation}
}}
Remarkably, Eqs. (\ref{waveqenv0}) and (\ref{poissons0}) describe, beyond the ponderomotive guiding center approximation \cite{1997PhPl....4..217M}, the spatio-temporal evolution of an e.m. pulse of arbitrary intensity, interacting with a Langmuire wave via nonlocal nonlinearities arising from relativistic effects. \textcolor{blue}{Although the PIC algorithms are now getting very fast, it is argued that fluid models for the plasma response may still be useful \cite{DreamBeams}, because the simulations of the next generation of LPA experiments (meter-scale, 10 GeV) will increase the computational requirements around 1000-fold.
The performance of LPA simulations can be vastly improved using the ponderomotive guiding center averaging procedure and modeling the envelope evolution of the laser field rather than the field itself \cite{2009AIPC.1086..309C}, but such procedure breaks down for very intense pump strengths. We overcome this limitation introducing a three-timescale procedure, and} seek the solution of Eq. (\ref{waveqenv0}) as the sum of its slowly and rapidly varying components, \textcolor{blue}{allowing the phase of the latter to vary on an intermediate time scale. We take} $\vec A_\bot = \vec A^{(0)}_\bot + \widetilde{\hspace{-1mm}\vec A}_\bot$, where $\vec A^{(0)}_\bot$ is the vector potential of the self-generated quasistationary magnetic field and $\widetilde{\hspace{-1mm}\vec A}_\bot$ is associated with the
electromagnetic wave of the laser,
\begin{equation}\label{modwav}
\widetilde{\hspace{-1mm}\vec A}_\bot = \vec A_{\bot_0}\left(t_2,\vec{r}_2\right) \,\, e^{i\left[\varphi\left(t_1,\vec{r}_1\right) - \omega' t + k'\left(z + ut \right)\right]} + c.c.
\end{equation}
The dimensionless frequency $\omega'$ and the wavenumber $k'$
are defined as
$\omega'={\omega}/{\omega_{pe}}$,
$k' = {ck}/{\omega_{pe}} = {\lambda_p}/{\lambda}$,
while $\omega$, $k$, and $\lambda$ are, respectively, the frequency, the wavenumber, and the wavelength of the electromagnetic wave propagating in an unperturbed plasma, that satisfy the linear dispersion relation $\omega = \sqrt{c^2 k^2 + \omega_{pe}^2}$. We drop the primes and write the dimensionless dispersion relation as
$\omega = \sqrt{k^2 + 1}$ .
The arguments of the laser phase $\varphi(t_1,\vec{r}_1)$ and amplitude $A_{\bot_0}(t_2,\vec{r}_2)$ in Eq. (\ref{modwav}) are given by
\begin{eqnarray}\label{othernot}
t_1 = \epsilon\, t - \epsilon^{-1} u z &,&
\vec{r}_1 = \vec{e}_x x + \vec{e}_y y + \epsilon^{-1}\vec{e}_z\, z, \\
\nonumber
t_2  = \epsilon^2 t_1= \epsilon^3 t - \epsilon\, u z &,&
\vec{r}_2 = \epsilon\, \vec{r}_1 = \epsilon \left(\vec e_x x + \vec e_y y\right) + \vec e_z z.
\end{eqnarray}
These are identified as the slow and intermediate scales, respectively, since under typical LPA conditions, with the plasma density $n_0 \leq 10^{19} \, {\rm cm}^{-3}$, we have $\omega\approx k > 12$ and the quantity $\epsilon \equiv 1/\omega \ll 1$ comprises a small parameter.

We adopt $u$ to be the group velocity of an e.m. wave,
$u = {d\omega}/{d k} = {k}/{\omega}$,
and introduce an auxiliary function $\kappa(\phi)$, localized and well-behaved, which is defined as
\begin{equation}\label{eqvarphi}
\kappa(\phi) = \left[\left(\nabla_1\varphi\right)^2 + 2\left({\partial\varphi}/{\partial t_1}\right) - \left({\partial\varphi}/{\partial t_1}\right)^2\right]^{1/2}.
\end{equation}
These permit us to rewrite Eqs. (\ref{waveqenv0}) and (\ref{poissons0}) as
\begin{eqnarray}
\nonumber
&2\, {\rm Re}\left\{e^{i\left[\varphi\left(t_1,\vec{r}_1\right)- \frac{t}{\omega} + k z\right]}\left[\alpha\, \vec{A}_{\bot_0} -
2\, i\, \epsilon^2 \left(1-\frac{\partial\varphi}{\partial t_1}\right)\frac{\partial \vec{A}_{\bot_0}}{\partial t_2}
\right.\right. \\
\nonumber
&-
\left.\left.
2\,i \, \epsilon \left(\nabla_1\varphi\cdot\nabla_2\right) \vec{A}_{\bot_0}  + \epsilon^4\,\frac{\partial^2 \vec{A}_{\bot_0}}{\partial t_2^2} - \epsilon^2\,\nabla_2^2 \vec{A}_{\bot_0}
\right]\right\} = \\
\nonumber
&\epsilon \left(\epsilon\,\frac{\partial}{\partial t_2}- u \, \frac{\partial}{\partial z_2}\right){\nabla_2}_\bot\phi -
\left(\epsilon^4\,\frac{\partial^2}{\partial t_2^2} - \epsilon^2\nabla_2^2 - \frac{1}{1-\phi}\right)\vec{A}^{(0)}_\bot \\
\label{waveqenv012} {}\\
& \left(\frac{\partial}{\partial z_2} - \epsilon\, u \, \frac{\partial}{\partial t_2}\right)^2\phi = \frac{\left(\phi -
1\right)^2 - 1 - {\vec{A}_\bot}^{\, 2}}{2 \left(\phi - 1\right)^2},
\label{poissons1}
\end{eqnarray}
where $\alpha = \kappa^2(\phi) + {\phi}/({1-\phi}) - i (\nabla_1^2\varphi - {\partial^2\varphi}/{\partial t_1^2})$ and
$\nabla_k = \vec{e}_x (\partial/\partial x_k) + \vec{e}_y (\partial/\partial y_k) +\vec{e}_z (\partial/\partial z_k)$, $k=1,2$.

The right-hand-side of the wave equation (\ref{waveqenv012}) is slowly varying in space and time and can not be resonant with the high-frequency oscillations on the left-hand-side. The slow vector potential, $A_\bot^{(0)}$, comes from the quasi stationary magnetic field generated in the laser-plasma interaction, for whose accurate description one needs to include also the kinetic effect that are responsible e.g. for the off-diagonal terms in the stress tensor for electrons \cite{Stress-magn-polje,Ja-i-Boba}, for the return electron current \cite{Ruski-rivju,Pegoraro-i-Califano,Askar-an}, etc.
As the derivation of our equations is based on a cold and unmagnetized plasma model \cite{mahajan,Shukla,2005PhPl...12c3101G}, they are valid only when the right-hand side of Eq. (\ref{waveqenv012}) is negligible. A scaling analysis of Eqs. (\ref{waveqenv012}), (\ref{poissons1}) shows that the slow vector potential $\vec{A}_\bot^{(0)}$ and the self-generated magnetic field can be neglected if
\begin{equation}\label{nomagneticfield}
{\rm max}(\phi, \, |\vec{A}_{\bot_0}|, \, |\vec{A}_\bot^{(0)}|) < (\epsilon \,\, \nabla_{2_\bot})^{-1} \, \partial/\partial z_2.
\end{equation}
As $\kappa(\phi)$ is a localized, well-behaved function of its argument the left-hand-side of Eq. (\ref{eqvarphi}) varies on the same spatial and temporal scales as the wake potential $\phi(t_2, \vec{r}_2)$, see Eq.s  (\ref{poissons0}) and (\ref{poissons1}). In other words, the function  $\kappa^2(\phi)$ is adopted to be a slowly varying function of the spatial variables $\vec{r}_1$ and a {\em very slowly} varying function of the variable $t_1$. Then, the fundamental solution for the phase $\varphi$ obtained from Eq. (\ref{eqvarphi}) is slowly varying with $t_1$, viz. $\partial\varphi/\partial t_1 \ll 1$. Using the new variables $\vec{\rho} = \vec{r}_1$ and $\tau = t_1 - \int^{\vec{r}_1}_{-\infty} {\vec{dl}\cdot\nabla_1\varphi} \,\, {\left(\nabla_1\varphi\right)^{-2}}$, Eq. (\ref{eqvarphi}) becomes
\begin{equation}\label{eqvarphi2}
\left(\nabla_\rho\varphi\right)^2 - \kappa^2\left(\phi\right) = {\cal O}\left(\epsilon^4\right),
\end{equation}
where derivatives with respect to the retarded time $\tau$ appear only in small terms, of order ${\cal O}(\epsilon^4)$. Consistently with the stationary, 1D approximation used in the derivation of the Poisson's equation (\ref{poissons0}), which is accurate to $\epsilon^2$, the right-hand-side of Eq. (\ref{eqvarphi2}) can be neglected.

Particularly simple is the case of a circularly polarized laser wave, $\vec A_{\bot_0} = (A_{\bot_0}/\sqrt{2})(\vec e_x + \vec e_y)$, when we have ${{\vec{A}_\bot}}^{\,\,2} = |A_{\bot_0}|^2$, i.e. the second harmonic is absent. Similarly, for a linearly polarized wave, we will neglect the second harmonic, whose contribution is nonresonant, and use  ${{\vec{A}_\bot}}^{\,\,2} \approx |A_{\bot_0}|^2$, where  $\vec A_{\bot_0} = A_{\bot_0}\,\vec e_x$.
Now, with the accuracy to $\epsilon^2$, our basic system of equations reduces to
\begin{eqnarray}
\nonumber
&
\left[\alpha_{Re}\left(\phi\right) + i \, \alpha_{Im}\left(\phi\right)\right] A_{\bot_0} -
2\, i\,\epsilon^2\,\, {\partial A_{\bot_0}}/{\partial t_2} -
\\
\label{poissons4}
&
2\,i\, \epsilon\left(\nabla_\rho\varphi\cdot\nabla_2\right) A_{\bot_0} - \epsilon^2\,\nabla_2^2 A_{\bot_0} = 0,
\\
\label{poissons04}
&
2 \,\, {\partial^2\phi}/{\partial z_2^2} = 1 - \left(1 + \left|A_{\bot_0}\right|^2\right)/\left(\phi - 1\right)^2,
\\
\label{eqvarphi4}
&
\left(\nabla_\rho\varphi\right)^2  = \kappa^2\left(\phi\right),
\end{eqnarray}
{where (\ref{eqvarphi4}) is an eikonal equation (the geometrical optics' limit), while} $\alpha_{Re}$ and $\alpha_{Im}$ are the real and imaginary parts of $\alpha$, respectively. With the accuracy to $\epsilon^2$ we have
\begin{equation}\label{alfaRe}
\alpha_{Re}\left(\phi\right) = {\phi}/({1-\phi}) + \kappa^2\left(\phi\right),
\quad
\alpha_{Im}\left(\phi\right) =  -\nabla_\rho^2\varphi.
\end{equation}
We adopt an auxiliary function $\kappa(\phi)$ which, asymptotically, in the \textcolor{blue}{MIR and SIR} reduce to the limits elucidated in \cite{moj_EPJD}. In the \textcolor{blue}{MIR}, $|\phi|\sim\epsilon^2$, a Schr\"{o}dinger equation with nonlocal cubic nonlinearity is to be recovered, which is realized within the scaling $\alpha_{Re}(\phi) \sim\kappa(\phi) \sim \varphi \sim {\cal O}(\epsilon^2)$, viz.
\textcolor{blue}{
\begin{equation}\label{waveqenvC}
\left[2i\,\omega\,\,{\partial}/{\partial t} +
\omega^{-2}\,{\partial^2}/{\partial z^2} + \nabla_\bot^2
- \phi\right]\, A_{\bot_0} = 0,
\end{equation}
\begin{equation}\label{poissons2C}
2\left({\partial^2}/{\partial z^2} + 1\right) \phi =
-{\left|A_{\bot_0}\right|^2}.
\end{equation}
}
In the ultrarelativistic regime \textcolor{blue}{(i.e. SIR)}, $1 \ll \phi \lesssim 1/\epsilon$ (for $\phi > 1/\epsilon \,$ one may not neglect the self-generated magnetic field), we have $\phi= - |A_{\bot_0}|$ and $|\nabla_\rho\varphi|  = 1$ and the wave equation describes an e.m. wave propagating in vacuum,
\begin{equation}\label{waveqenvBF}
\textcolor{blue}{
\left(2i\,\omega\,\,{\partial}/{\partial t} + \nabla_\bot^2\right)\, A_{\bot_0}
= -A_{\bot_0}/\phi \,\, \to 0,
}
\end{equation}
\textcolor{blue}{while $\phi$ is found from $2\,\,{\partial^2\phi}/{\partial z^2} =
1- \left|A_{\bot_0}\right|^2/\phi^2$. These are} realized when $\alpha_{Re}\to 1/\phi\to - \epsilon$  and  $\kappa\to 1$.
Close to the edges of an ultrarelativistic laser pulse, in the region where $\phi\sim {\cal O}(1)$, $\alpha_{Re}(\phi)$ needs to be sufficiently small, so that the nonlinear term $\alpha A_{\bot_0}$ has the same scaling as the linear terms in the wave equation.
We adopt a simple expression
$\alpha_{Re}(\phi) = \phi(1 + \phi^2)/(1 - \phi)^4$, and consequently:
\begin{equation}\label{choice2}
\kappa(\phi) = -\left[{\phi}/\left({1 - \phi}\right)\right]\left[1+{2}/{\left(1 - \phi\right)^2}\right]^\frac{1}{2}.
\end{equation}

\section{Numericalal results}

\textcolor{blue}{For an efficient LPA, one needs to find, within the technical constraints of the available lasers, the parameters that optimize the system's performance. To initiate such optimization, first we make an 'educated guess' about the most stable lengthscale of the laser pulse, by finding an analytic stationary solution of Eqs. (\ref{poissons4})-(\ref{eqvarphi4}) and (\ref{choice2}) under simplified, albeit unphysical, conditions.} For this, we neglect all transverse effects, $\partial/\partial x = \partial/\partial y =0$, and the nonlocality effects in the Poisson's equation (\ref{poissons04}), viz. $({\partial}/{\partial z_2} - \epsilon\, u \, {\partial}/{\partial t_2})^2\phi \to 0$. The wave equation is then readily solved in the form $\vec{A}_{\bot_0} = a_L(t_2,z_2) \, \exp\{i[\delta k \,  z_2 - \delta\omega \, t_2 - \varphi(t_1,z_1)]\}$, where $a_L$, $\delta k$ and $\delta\omega$ are purely real quantities and the subscript $L$ denotes a local solution. Separating real and imaginary parts of Eq. (\ref{poissons4}), and after some algebra, we have
\begin{equation}\label{local}
a_L^{-1}({\partial^2 a_L}/{\partial z_2^2}) + \epsilon^{-2}\left[1-\left(1 + a_L^2\right)^{-1/2}\right] = \delta k^2,
\end{equation}
which is easily integrated by quadratures
\begin{equation}\label{MadelungL7}
z_2 - \delta k \, t_2 =\left({\epsilon}/{\sqrt{2}}\right)\int {d a_L} \,{\left[c_1 a_L^2 + \left(1 + a_L^2\right)^{1/2}-1\right]^{-1/2}}.
\end{equation}
Here $c_1\equiv(\epsilon^2/2)(\delta k^2 - 2\,\delta\omega - 1/\epsilon^2)$ is arbitrary constant and the maximum value of $a_L$ is $a_{max} = (1 /c_1)\sqrt{1 + 2 c_1}$. A numerical integration of Eq. (\ref{MadelungL7}), with a typical LPA experimental value $\epsilon = 1/12$ and with the maximum laser vector potential $a_{max} = 14.12$, which is well within the ultrarelativistic regime determined by $a \geq 4$ \cite{2006PhPl...13e6709L}, reveals a typical
bell-shaped profile. Its e-folding length $L_{\Vert ef}$, i.e. the separation between the points where the intensity is reduced by the factor $e$, $a_L^2(L_{\Vert ef}/2) = a_{max}^2/e$, is given by $L_{\Vert ef} = 0.65$, or $L_{\Vert ef} = 1.1 \, \mu{\rm m}$ in non-scaled variables.

\textcolor{blue}{In the next step, we study the influence of the finite width and of the nonlocal effects on the pulse evolution.} We solve numerically Eqs. (\ref{poissons4})-(\ref{eqvarphi4}) and (\ref{choice2}) in 2D (with $\partial/\partial y_2$ $= 0$), in an initially quiescent plasma $\phi\left(x_2,z_2,0\right)$ $= \varphi\left(x_2, z_2, 0\right) = 0$. The initial shape of the laser pulse was adopted \textcolor{blue}{in the form of the stationary local solution} (\ref{MadelungL7}), but with a Gaussian transverse profile with the width $L_x$, $A_{\bot_0}(x_2, z_2, 0)$ $= a_L(z_2/L_z)$ $\exp({i\,\delta k\, z_2})$ $\exp{(-x_2^2/2 L_x^2)}$. Here $a_L$ is given by (\ref{MadelungL7}) and we take $\delta k = -0.5$, $L_x = 7.5$, and $L_z = 0.8$. Such laser pulse has the initial root mean square width (in non-scaled variables) $L_{\bot rms} =  150 \, \mu{\rm m}$, the laser energy $E \approx 125 \, {\rm J}$ and the power $P\approx 50 \, {\rm PW}$. The solution, \textcolor{blue}{displayed in Figs. 1--3}, was followed up to $t_{2_{max}}\gtrsim 1$. \textcolor{blue}{During such short time} no transverse contraction was observed. The folding of the pancake pulse to a V-shape, known in the \textcolor{blue}{MIR} \cite{moj_EPJD,2005PhPl...12c3101G} was not observed either. Due to the stretching in the forward direction, the laser amplitude rapidly dropped,  to $|A_{max}| \sim 5$, which is still within the ultrarelativistic regime, $|A_{\bot_0}| \geq 4$. By the time $t_{2_{max}}$ the maximum forward stretching was reached. It occurred mostly in the central part of the pulse, where the amplitude was the largest.
Inside the pulse, almost all electrons had been pushed out by the ponderomotive force and the laser light practically propagated in a vacuum, i.e. the core of the pulse propagated with the speed of light and the nonlinear effects were weak. The nonlocality produced an effective mixing with the front edge of the pulse (that tends to propagate with the group velocity), which is then \textcolor{blue}{pushed} forward by the core. An electrostatic wake with a rather large first potential minimum $|\phi_{max}|$ $\gtrsim 1.5$ developed roughly during the time $t_2 = 0.5$ and grew steadily until the maximum time was reached.
The nonlinear phase $\varphi$ emerged simultaneously with the wake, producing a substantial bending of the laser wave front.

\begin{figure}[h]
\centering
\includegraphics[width=85mm]{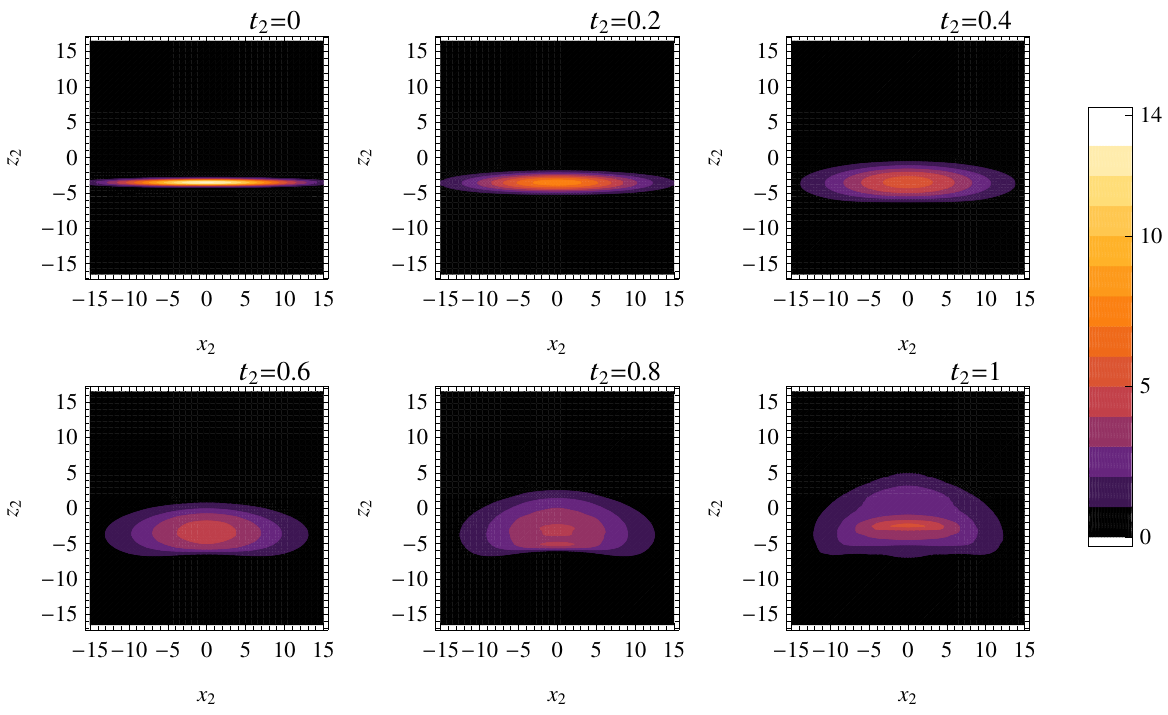}
\caption{Evolution, in an ultrarelativistic regime, of the envelope $|A_{\bot_0}(x_2, z_2, t_2)|$ of a pancake-shaped laser pulse. The initial condition is $A_{\bot_0}(x_2, z_2, 0) = a_L(z_2/L_z)\,\, \exp{(-x_2^2/2 L_x^2)}$ $\exp({i\,\delta k\, z_2})$, with $\delta k = -0.5$, $L_z = 0.8$, and $L_x = 7.5$ and $\phi\left(x_2,z_2,0\right) = \varphi\left(x_2,z_2,0\right) = 0$. In the non-scaled variables, the initial pulse length and width are $\sim0.9 \, \mu{\rm m}$ and  $\sim 150 \, \mu{\rm m}$. Dimensionless time $t_{2_{max}} = 1$ corresponds to $9.69\times 10^{-12} \,{\rm s}$, during which time the pulse travels $2.9 \, $mm. (color online). } \label{sir_laser}
\end{figure}
\begin{figure}[h]
\centering
\includegraphics[width=85mm]{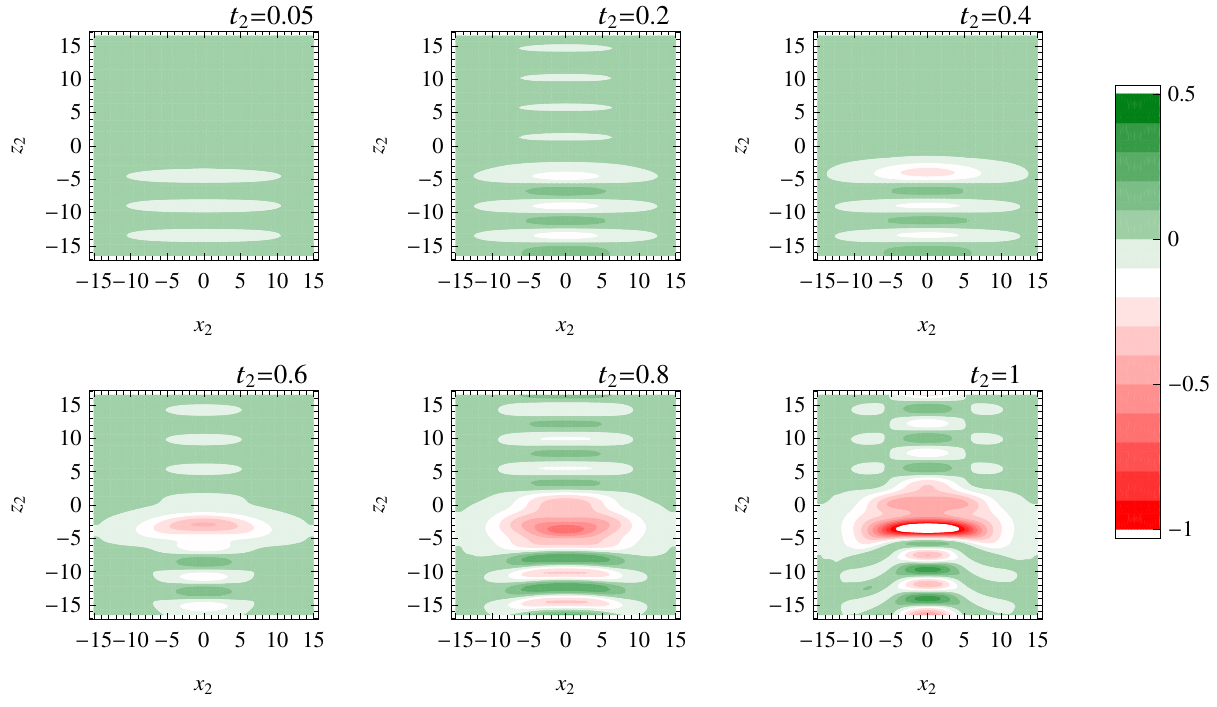}
\caption{The electrostatic wake potential $\phi(x_2, z_2, t_2)$, produced by the laser pulse displayed in Fig. \ref{sir_laser}.  (color online) } \label{sir_wake}
\end{figure}
\begin{figure}[h]
\includegraphics[width=85mm]{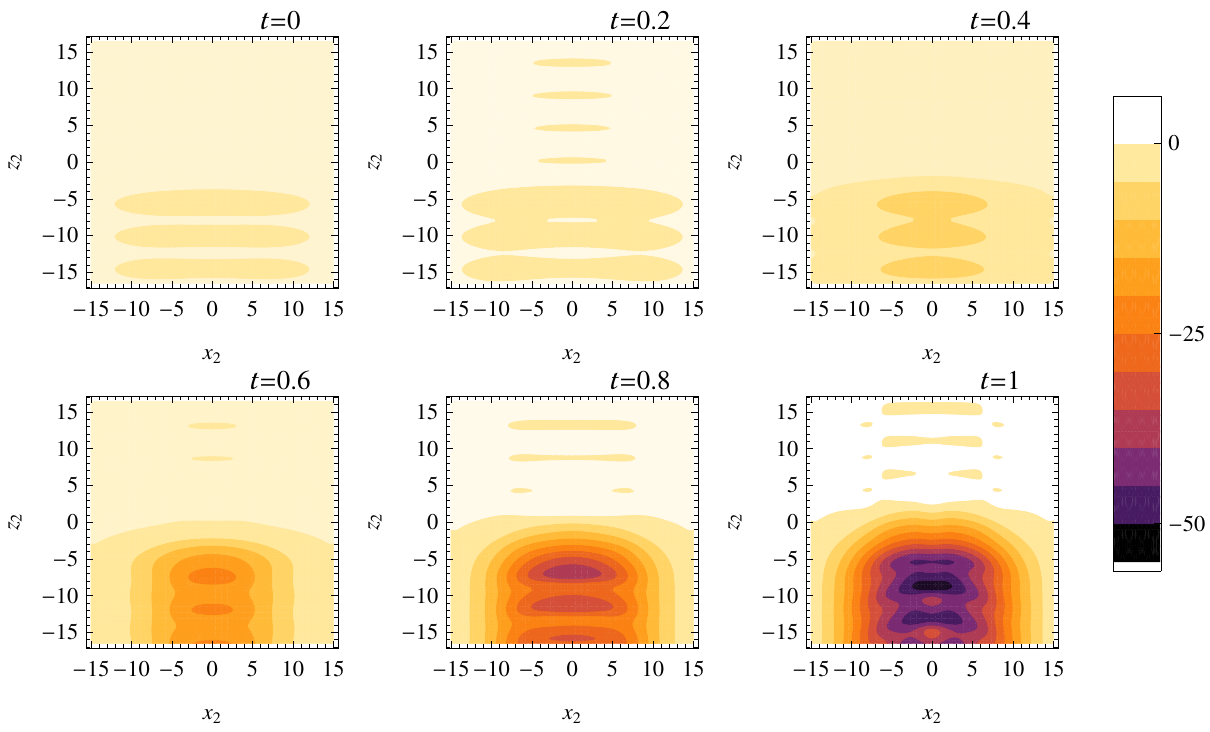}
\caption{The nonlinear phase $\varphi(x_2, z_2, t_2)$ of the laser pulse displayed in Fig. \ref{sir_laser}. (color online)} \label{sir_NLPhase}
\end{figure}

\section{Conclusions}

In this letter we have studied, using a semi-analytic hydrodynamic description, a \textcolor{blue}{SIR} regime of the propagation of pancake-shaped laser pulses through an unmagnetized plasma, with specifications envisaged for the next generation of LPA experiments. We derive novel model equations, based on a three-timescale description, that account for the evolution of the nonlinear phase of the laser wave. At very large laser intensities this gives a smooth transition to a nondispersive e.m. wave and the saturation of the nonlocal nonlinearity. These equations are solved numerically in the regime when \textcolor{blue}{the ultrarelativistic} electrons are almost expelled by the radiation pressure of a femtosecond laser pulse focussed to a $\gtrsim 100 \, \mu{\rm m}$ spot. We could follow the pulse during its travel along a $2.9 \, $mm long path, which coincides with the dimensions of the He plasma in the self-injection LPA experiment \cite{2009RvMP...81.1229E}. Practically no transverse self-focussing \textcolor{blue}{and fillamentation} have been observed for such broad pulse, but its Raleigh length is sufficiently long to allow for an efficient electron acceleration without self-guiding. \textcolor{blue}{The plasma wake, whose peak potential is \textcolor{blue}{$\phi \gtrsim 1.5$}, preserves its length ($\sim \lambda_p$), while the nonlocality effects stretch the laser pulse almost tenfold in the forward direction. A moderate stretching has been known in the \textit{mildly relativistic} \textcolor{blue}{(MIR)} regime \cite{moj_EPJD,2005PhPl...12c3101G}, arising from the dispersion of the laser pulses if, initially, they were sufficiently shorter than the plasma length. Pulses that are $\gtrsim\lambda_p$, undergo longitudinal compression into a "laser piston"
\cite{2012JPlPh..78..469C}. For them, a mild stretching may be desirable \cite{tamne_struje}, because it compensates the nonlinear red-shift and delays the formation of the "piston", which reduces the dark current.}

Our calculations have been performed for a laser with the energy $\sim 125 \, {\rm J}$ per pulse and the duration $T\lesssim 10 \, {\rm fs}$, providing the power of several tens of petawatts and the intensity $I \sim 10^{20} \, {\rm W/cm}^2$.
Lasers with such energy specifications are planned for the near future, mostly in the Nd:Glass technology. The Vulcan upgrade \cite{2007SPIE.6735E..17C,TUP083} will have a new  laser beamline with $300 \, {\rm J}$ in $30 \, {\rm fs}$ ($10 \, {\rm PW}$), that can be focussed to $10^{23} \, {\rm W/cm}^2$. Extreme Light Infrastructure (ELI) \cite{TUPB57} will produce in its second section, planned for a later phase, a $10 \, {\rm PW}$ beamline compressed to $130 \, {\rm fs}$, providing an on-target power density $I > 10^{23} \, {\rm W/cm}^2$. The electron acceleration to $2-5 \, {\rm GeV}$ is expected to be reached by 2019 and up to $50 \, {\rm GeV}$ after 2020. Ultra-strong Ti:Sapphire lasers have also been planned, such as Astra-Gemini \cite{2007SPIE.6735E..17C}, a dual beam upgrade to a PW class of the existing Astra facility that will supply $10^{22} \, {\rm W/cm}^2$ on target. Each beam will have $15 \, {\rm J}$ compressed to $30 \, {\rm fs}$, supplying $0.5 \, {\rm PW}$. \textcolor{blue}{Further compression to $\sim 5 \, {\rm fs}$ ($\sim 2$ oscillations), is possible by the use of photon deceleration or thin plasma lenses \cite{single_cycle_Tsung_Mori,2001PhRvE..63b6411R}, and the transverse filamentation can be stabilized by a periodic plasma-vacuum structure \cite{pukhov-compress}.}
Our semi-analytic fluid theory may be a valuable tool for the predictions and the analyses of LPA experiments in the ultrarelativistic regime with these lasers, focussed to a spot $\gtrsim 100 \, \mu{\rm m}$, \textcolor{blue}{for which it can provide an estimate for the accelerating wakefield and its dynamics.
While the oversimplified local model preferred very short (single oscillation) laser pulses, the observed stretching implies that an optimized LPA system may require a longer pulse. Kinetic effects, such as the plasma wave-breaking, the trapping of resonant particles and their subsequent acceleration, is not included in the present analysis. They are the subject of our study in progress, to be presented later.}

\begin{acknowledgements}This work was supported in part by the Serbian MPNTR grant 171006. D.J. acknowledges financial support from the FAI
of the Italian INFN and the hospitality of Dipartimento di Fisica, Universit\`{a} di Napoli "Federico II".
\end{acknowledgements}

\bibliographystyle{eplbib}

\begin{thebibliography}{10}
\expandafter\ifx\csname url\endcsname\relax\def\url#1{\texttt{#1}}\fi
\bibitem{LWF2}
\Name{{Tajima} T. \and {Dawson} J.~M.} \REVIEW{Phys. Rev. Lett.}{43}{1979}{267}.
\bibitem{2009RvMP...81.1229E}
\Name{{Esarey} E., {Schroeder} C.~B. \and {Leemans} W.~P.} \REVIEW{Rev. Mod. Phys.}{81}{2009}{1229}.
\bibitem{2011PhyU...54....9K}
\Name{{Korzhimanov} A.~V., {Gonoskov} A.~A., {Khazanov} E.~A. \and {Sergeev} A.~M.} \REVIEW{Phys. Usp.}{54}{2011}{9}.
\bibitem{2013NatCo...4E1988W}
\Name{{Wang} X., {Zgadzaj} R., {Fazel} N. et al} \REVIEW{Nature Comm.}{4}{2013}{}.
\bibitem{moj_EPJD}
\Name{{Jovanovi{\'c}} D., {Fedele} R., {Tanjia} F., {De Nicola} S. \and {Gizzi} L.~A.} \REVIEW{Eur. Phys. J. D}{66}{2012}{328}.
\bibitem{2006PhPl...13e6709L}
\Name{{Lu} W., {Huang} C., {Zhou} M. et al} \REVIEW{Phys. Plas.}{13}{2006}{056709}.
\textcolor{blue}{
\bibitem{2004PhPl...11.5256K}
\Name{{Kostyukov} I., {Pukhov} A. \and {Kiselev} S.} \REVIEW{Phys. Plas.}{11}{2004}{5256}.}
\textcolor{blue}{
\bibitem{2002ApPhB..74..355P}
\Name{{Pukhov} A. \and {Meyer-ter-Vehn} J.} \REVIEW{Appl. Phys. B}{74}{2002}{355}.}
\bibitem{2005PhPl...12d3109G}
\Name{{Gordienko} S. \and {Pukhov} A.} \REVIEW{Phys. Plas.}{12}{2005}{043109}.
\bibitem{PukhovBubbles}
\Name{{Pukhov} A. \and {Gordienko} S.} \REVIEW{Phil. Trans. R. Soc. A}{364}{2006}{623}.
\bibitem{2005PhPl...12c3101G}
\Name{{Gorbunov} L.~M., {Kalmykov} S.~Y. \and {Mora} P.} \REVIEW{Phys. Plas.}{12}{2005}{033101}.
\textcolor{blue}{
\bibitem{1997PhRvL..78..879B}
\Name{{Borghesi} M., {MacKinnon} A.~J., {Barringer} L. et al} \REVIEW{Phys. Rev. Lett.}{78}{1997}{879}.}
\textcolor{blue}{
\bibitem{1997IJQE...33.1879E}
\Name{{Esarey} E., {Sprangle} P., {Krall} J. \and {Ting} A.} \REVIEW{IEEE J. Quant. Elec.}{33}{1997}{1879}.}
\textcolor{blue}{
\bibitem{2001AIPC..569..214E}
\Name{{Esarey} E., {Shadwick} B.~A., {Schroeder} C.~B. et al} \Book{AIP  Conf. Ser.} Vol. 569, 2001 pp. 214--222.}
\bibitem{Gizzi2013}
\Name{{Gizzi} L.~A., {Anania} M.~P., {Gatti} G. et al} \REVIEW{Nucl. Inst. Met. Phys. Res. B}{309}{2013}{202–209}.
\bibitem{sprangle}
\Name{{Sprangle} P., {Esarey} E. \and {Ting} A.} \REVIEW{Phys. Rev. Lett.}{64}{1990}{2011}.
\bibitem{mahajan}
\Name{{Berezhiani} V.~I. \and {Mahajan} S.~M.} \REVIEW{Phys. Rev. Lett.}{73}{1994}{1837}.
\bibitem{Shukla}
\Name{{Sharma} A., {Kourakis} I. \and {Shukla} P.~K.} \REVIEW{Phys. Rev. E}{82}{2010}{016402}.
\textcolor{blue}{
\bibitem{2011PhRvL.106m5002S}
\Name{{Schroeder} C.~B., {Benedetti} C., {Esarey} E. \and {Leemans} W.~P.} \REVIEW{Phys. Rev. Lett.}{106}{2011}{135002}.}
\textcolor{blue}{
\bibitem{2010PhPl...17e4704K}
\Name{{Kostyukov} I. \and {Pukhov} A.} \REVIEW{Phys. Plas.}{17}{2010}{054704}.}
\bibitem{1997PhPl....4..217M}
\Name{{Mora} P. \and {Antonsen}, Jr. T.~M.} \REVIEW{Phys. Plas.}{4}{1997}{217}.
\bibitem{DreamBeams}
\Name{{Mori} W., {An} W., {Decyk} V.~K. et al} proc. of \Book{SciDAC, Chattanooga, Tennessee, U.S.A.} 2010 pp. 261--276.
\textcolor{blue}{
\bibitem{2009AIPC.1086..309C}
\Name{{Cowan} B., {Bruhwiler} D., {Cormier-Michel} E. et al} \Book{AIP  Conf. Ser.}, edited by \Name{{Schroeder} C.~B.,
  {Leemans} W. \and {Esarey} E.} Vol. 1086, 2009 pp. 309--314.}
\bibitem{Stress-magn-polje}
\Name{{Giulietti} A., {Tomassini} P., {Galimberti} M. et al} \REVIEW{Phys. Plas.}{13}{2006}{093103}.
\bibitem{Ja-i-Boba}
\Name{{Jovanovic} D. \and {Vukovi{\'c}} S.} \REVIEW{Physica B+C}{125}{1984}{369}.
\bibitem{Ruski-rivju}
\Name{{Belyaev} V.~S., {Krainov} V.~P., {Lisitsa} V.~S. \and {Matafonov} A.~P.} \REVIEW{Phys. Usp.}{51}{2008}{793}.
\bibitem{Pegoraro-i-Califano}
\Name{{Pegoraro} F., {Bulanov} S.~V., {Califano} F. \and {Lontano} M.} \REVIEW{Phys. Scr. T}{63}{1996}{262}.
\bibitem{Askar-an}
\Name{{Askar'yan} G.~A., {Bulanov} S.~V., {Pegoraro} F. \and {Pukhov} A.~M.} \REVIEW{Plas. Phys. Rep.}{21}{1995}{835}.
\bibitem{2012JPlPh..78..469C}
\Name{{Cowan} B.~M., {Kalmykov} S.~Y., {Beck} A. et al} \REVIEW{J. Plasma Phys.}{78}{2012}{469}.
\textcolor{blue}{\bibitem{tamne_struje}
\Name{{Kalmykov} S.~Y., {Shadwick} B.~A., {Beck} A. \and {Lefebvre} E.} proc. of \Book{InTech 2011, Rijeka, Croatia}, edited by \Name{{Andreev} A.~V.}
  2011 pp. 113--138.}
\bibitem{2007SPIE.6735E..17C}
\Name{{Chekhlov} O., {Divall} E.~J., {Ertel} K. et al} \Book{SPIE Conf. Ser.} Vol. 6735 2007 pp. 67350J--1--67350J--7.
\bibitem{TUP083}
\Name{{Pepler} D., {Boyle} A., {Collier} J. et al} proc. of \Book{ICALEPCS2009, Kobe, Japan} 2009 pp. 272--274.
\bibitem{TUPB57}
\Name{{Pribyl} L., {Juha} L., {Korn} G. et al}
proc. of \Book{IBIC2012, Tsukuba, Japan} 2012 pp. 482--485.
\textcolor{blue}{\bibitem{single_cycle_Tsung_Mori}
\Name{{Tsung} F.~S., {Ren} C., {Silva} L.~O. et al} \REVIEW{Proc. Nat. Acad. Sci. USA}{99}{2002}{29–32}.}
\textcolor{blue}{
\bibitem{2001PhRvE..63b6411R}
\Name{{Ren} C., {Duda} B.~J., {Hemker} R.~G. et al} \REVIEW{Phys. Rev. E}{63}{2001}{026411}.}
\textcolor{blue}{
\bibitem{pukhov-compress}
\Name{{Shorokhov} O., {Pukhov} A. \and {Kostyukov} I.} \REVIEW{Phys. Rev. Lett.}{91}{2003}{265002}.}
\end{thebibliography}

\end{document}